\documentclass{epl}

\title{ESR theory for interacting 1D quantum wires}
\author{A. De Martino\inst{1,2} \and R. Egger\inst{1}}
\institute{
  \inst{1} Institut f\"ur Theoretische Physik, Heinrich-Heine-Universit\"at,
D-40225 D\"usseldorf\\
  \inst{2} Fakult\"at f\"ur Physik, Albert-Ludwigs-Universit\"at,
D-79104 Freiburg
}

\begin{document}

\maketitle

\begin{abstract}
We compute the electron spin resonance (ESR) intensity 
for one-dimensional quantum wires in semiconductor heterostructures,
taking into account electron-electron interactions
and spin-orbit coupling.  The ESR spectrum is shown to be
very sensitive to interactions.  While in the absence of
interactions, the spectrum is a flat band, characteristic
threshold singularities appear in the interacting limit.
This suggests the practical use of ESR to reveal
spin dynamics in a Luttinger liquid.
\end{abstract}

The search for the Luttinger liquid (LL) behavior
of one-dimensional (1D) interacting fermions 
\cite{tomonaga,luttinger,haldane,gogolin} has
a rather long history.   Recent experimental work has reported
strong evidence via the tunneling density of
states in single-wall carbon nanotubes \cite{bockrath,yao},
for the edge states in the fractional quantum Hall
effect \cite{chang} (which  supposedly are chiral LLs
not further discussed here), and via the conductance \cite{tarucha}
and resonant tunneling \cite{auslaender}
in a quantum wire (QW) in semiconductor heterostructures.
These studies have so far only probed the charge dynamics 
and are intrinsically insensitive 
to {\sl spin dynamics}.  However, spin dynamics can reveal some of
the most interesting and spectacular aspects of the
non-Fermi liquid nature of a LL, in particular the
phenomena of spin-charge separation and electron fractionalization 
\cite{gogolin}.
In this Letter, we suggest to probe the spin dynamics of 
a LL via electron
spin resonance (ESR), and provide detailed theoretical predictions 
for the ESR spectrum in an interacting 1D QW.  
ESR spectra have already been obtained experimentally for the 2D electron
gas in heterostructures \cite{exp}, and new fabrication
advances such as cleaved-edge overgrowth techniques \cite{yacoby} indicate
that such experiments could indeed be done for QWs as well.

The ESR spectrum of a LL turns out to be extremely sensitive to the
presence and precise nature of the spin-orbit coupling mechanism.
This is intuitively clear as spin-orbit coupling can sometimes destroy
spin-charge separation, and nearly always represents the leading term
breaking $SU(2)$ invariance.
Unless this symmetry is broken, the ESR spectrum (including
electron-electron interactions)
is simply a delta peak at the Zeeman energy.
Remarkably, the spin-orbit coupling 
in semiconductor QWs will spoil spin-charge separation
according to refs.~\cite{moroz1,moroz2},
while this does not happen in nanotubes \cite{later}. 
Therefore ESR theory for other realizations
of the LL (such as nanotubes) is quite different 
and will be given elsewhere \cite{later}.  
Furthermore, the ESR theory of 1D spin chains 
 \cite{affleck} is also distinct due to the presence
of Dzyaloshinskii-Moriya interactions and the 
absence of charge degrees of freedom. 
To be specific, here we focus on the practically important
case of QWs.  For the theoretical description of
the QW including electron-electron interactions, 
we use the LL model.

The standard LL Hamiltonian is expressed in
terms of charge and spin bosons $\phi_\nu$ 
 ($\nu=\rho,\sigma$) and their conjugate momenta $\Pi_\nu$
 \cite{gogolin},
\begin{equation} \label{llham}
H = \sum_{\nu=\rho,\sigma} 
 \frac{v_\nu}{2} \int dx \left[   K_\nu  \Pi_\nu^2 + 
K^{-1}_\nu (\partial_x \phi_\nu)^2 \right] ~,
\end{equation}
with respective effective velocities 
$v_\nu$ and LL parameters $K_\nu$. 
Under Abelian bosonization,
the right- and left-moving $(r=R,L=\pm$) electron operator 
for spin $\alpha= \uparrow, \downarrow=\pm$ is
\begin{equation} \label{fermop}
\psi_{r\alpha}(x) = \frac{1}{\sqrt{2\pi \Lambda}} \exp\left\{ i\sqrt{\pi/2} 
[ r\phi_\rho + r \alpha \phi_\sigma + \theta_\rho + 
\alpha \theta_\sigma ]  \right \} ~,
\end{equation}
where $\theta_\nu = - \int_{-\infty}^x dx' \Pi_\nu$
and $\Lambda$ is a short-distance cutoff-length.
To ensure $SU(2)$ invariance of eq.~(\ref{llham}), we put 
$K_\sigma=1$, while for repulsive interactions,
the charge interaction parameter $K_\rho\equiv g <1$.
In addition, to respect Galilean invariance 
manifest in QWs, the velocities $v_\nu$ are taken as
$v_\sigma=v_F$ and $v_\rho=v_F/g$ with the Fermi velocity $v_F$.
To simplify notation, below we often put $v_F=\hbar=1$.

The main source of spin-orbit coupling in QWs  
comes from the Rashba and confinement electric fields, 
and for sufficiently large fields leads to a deformation of 
each branch of the single-particle dispersion relation \cite{moroz1}.  
The net result of this deformation is the breaking of chiral symmetry, since 
the Fermi velocities for right- and left-moving electrons with the
same spin will become different.
Therefore, with $v_{R\uparrow}=v_{L\downarrow}=v_1$ and
$v_{R\downarrow}=v_{L\uparrow}=v_2$, we have $\delta v = v_2-v_1 > 0$
and $v_1 + v_2 = 2 v_F$.  
A dimensionless measure of the spin-orbit coupling strength 
is then given by $\lambda \equiv \delta v/2 v_F \ll 1$.
Concrete estimates for quantum wires yield values of the order
$\lambda \approx 0.1$ \cite{moroz2}.
In effect, the spin-orbit interaction produces an additional $SU(2)$
symmetry-breaking term that can be expressed as 
\begin{equation} \label{so}
H'  = \lambda \int dx \left[ \Pi_\rho  \partial_x \phi_\sigma 
+ \Pi_\sigma  \partial_x \phi_\rho \right] ~.
\end{equation}
Under an external static magnetic field $B$ along the $z$ axis, 
we have to add a Zeeman term, 
\begin{equation} \label{zeeman} 
H_Z = - B \int dx \, S^z(x) =  -\frac{B}{\sqrt{2\pi}}
 \int dx  \,\partial_x \phi_\sigma \;,
\end{equation}
with $g_e \mu_B = 1$ for simplicity. 
Due to band curvature, the Zeeman term also gives rise to a small splitting of 
the spin-up and spin-down Fermi velocities \cite{kimura},
$\lambda_\sigma = |v_{F\uparrow}-v_{F\downarrow}|/2v_F \approx
 B/ 2E_F$.  For the most interesting cleaved-edge
overgrowth samples, the 1D Fermi level is $E_F>20$~meV \cite{yacoby},
 and therefore $\lambda_\sigma \ll \lambda$  even for strong magnetic
fields.  Based on this observation, the Zeeman-induced
splitting of the Fermi velocities is neglected here.

The ESR intensity at frequency $\omega$ 
is then proportional to the Fourier-transformed transverse spin 
correlation function, 
\begin{equation} \label{ESR}
I(\omega) = \int \! dt dx  \, e^{i\omega t} 
\langle S^+(t,x) S^-(0,0) \rangle ~,
\end{equation}
where $S^{\pm}(t,x)$ is the transverse component of the uniform part 
of the spin operator, $S^{\pm } = J^{\pm}_R +J^{\pm}_L$, 
with the spin currents
$\vec J_{R,L} = \psi^\dagger_{R,L} (\vec \sigma/2) \psi_{R,L}$.
The Zeeman term (\ref{zeeman}) 
can now be transformed away by shifting the field 
$\phi_\sigma$ and 
simultaneously $\Pi_\rho$, 
\begin{equation} \label{shifts}
\phi_\sigma  \to  \phi_\sigma + 
\frac{Bx}{\sqrt{2\pi}} 
\frac{1}{1-\lambda^2} ~, \quad 
\Pi_\rho     \to  \Pi_\rho - 
\frac{B}{\sqrt{2\pi}} \frac{\lambda}{1-\lambda^2} ~.
\end{equation}
This is a direct consequence of the fact that the spin-orbit term
(\ref{so}) spoils spin-charge separation. 
Albeit the magnetic field only couples to the spin, 
it now also affects the charge sector, and the combined
shift (\ref{shifts}) is necessary to transform away the
Zeeman term.
Of course, the fermion operators and the transverse components of the 
uniform part of spin operator are then modified,
\[
\psi_{r\alpha}(x)  \to 
\exp\left \{i r\alpha Bx \frac{1+r\alpha\lambda}{2(1-\lambda^2)}
\right\} \psi_{r\alpha}(x) \;,
\]
and therefore 
\begin{equation} \label{s+}
J^\pm_R (x)  \to  e^{\mp i k_\sigma x} J^{\pm}_R (x) ~, \quad
J^\pm_L (x) \to e^{\pm i k_\sigma x} J^{\pm}_L (x) ~,
\quad 
v_F k_\sigma = B/(1-\lambda^2) ~.
\end{equation}

Since the Hamiltonian remains quadratic in the boson fields,
the full problem can be solved exactly. 
We first focus on the $T=0$ limit where interaction
effects are most pronounced.
After straightforward but tedious algebra, the
fermion propagator and hence the transverse
spin correlation function are obtained.  Using 
eqs.~(\ref{fermop}) and (\ref{s+}), 
\begin{equation} \label{jrjr}
\langle J^+_R(t,x) J^-_R(0,0) \rangle = 
\frac{e^{ - i k_\sigma x }}{(2\pi \Lambda)^2} 
\prod_{i=1,2} \left( \frac{ i \Lambda}{x - u_i t + i \Lambda}
\right)^{\xi^+_i} \left( \frac{ - i \Lambda}{x + u_i t - i \Lambda}
\right)^{\xi^-_i} ~, 
\end{equation}
with the same expression but $x\to -x$ for 
the left-moving part.
Here the eigenmode velocities $u_1$ and $u_2$ are in units of $v_F$:
\begin{equation}
u_{1,2} = \frac{1}{\sqrt{2}} \left[ g^{-2}  + 1 + 
2 \lambda^2  \mp \sqrt{ (g^{-2} - 1)^2 +8 \lambda^2 
(g^{-2} + 1)}  \right]^{1/2} ~.
\end{equation}
In the absence of spin-orbit coupling, $\lambda = 0$, these velocities reduce 
to the spin and charge velocities of the LL,  
$u_{1} \to v_F$ and $u_{2} \to v_F/g$, while
in the non-interacting case, $g=1$, we recover
$u_1\to v_1$ and $u_2 \to v_2$.
The exponents $\xi_i^{\pm}$ in eq.~(\ref{jrjr}) are given by
\begin{equation} \label{exponents}
\xi^\pm_i  =  (-1)^i  \frac{  u_i^2 -g^{-2} + (g^{-2}+1) \lambda^2/2
\pm u_i (u_i^2 - g^{-2}-\lambda^2) }
{ u_i (u_2^2-u_1^2) }  ~.
\end{equation}
The ESR intensity (\ref{ESR}) follows from eq.~(\ref{jrjr})
and the corresponding left-moving part:
\begin{equation} \label{esrfinal}
I(\omega) = 2 \, {\rm Re}\, \int dx dt 
\frac{e^{i\omega t - i k_\sigma x }}{(2\pi \Lambda)^2} 
\prod_{i=1,2} \left( \frac{ i \Lambda}{x - u_i t + i \Lambda}
\right)^{\xi^+_i} \left( \frac{ - i \Lambda}{x + u_i t - i \Lambda}
\right)^{\xi^-_i} ~. 
\end{equation}
We are unable to compute this double integral in
closed analytical form for arbitrary parameters.
However, it has exactly the same structure as the integrals 
appearing in the computation of the spectral function for
the spinful LL \cite{meden,meden2}. 
As in that case, one can determine the thresholds 
and the singularities of the ESR spectrum. 

Before that let us briefly analyze some limiting cases.
In the {\sl non-interacting} limit $g=1$, the exponents (\ref{exponents})
simplify to $\xi^+_{1,2} = 1$ and $\xi^-_{1,2} = 0$. 
With these values for the exponents 
the double integral in eq.~(\ref{ESR}) can easily be 
evaluated, and the ESR intensity is uniform
over a finite range of frequencies with bandwidth
$\Delta\omega =  2B\lambda/(1 - \lambda^2)$ 
and zero otherwise,
\begin{equation} \label{Inil}
I(\omega) = \frac{\theta ( v_2 k_\sigma   -\omega ) \, \theta (
\omega - v_1 k_\sigma )}{v_F \lambda} ~.
\end{equation}
Clearly, in the limit of vanishing spin-orbit coupling
($\lambda \to 0$), the single delta peak at $\omega=B$
is recovered. The integrated spectrum is given by
\begin{equation} \label{intsp}
\int \! \frac{d\omega}{2\pi} \, I(\omega) = \frac{1}{\pi}
\frac{B}{v_F}\frac{1}{1 - \lambda^2} ~.
\end{equation}
Furthermore, in the absence of spin-orbit coupling but with
$g<1$, it is straightforward to recover the delta peak
at $\omega=B$ directly from eq.~(\ref{esrfinal}).  

Let us now turn to the general  case with $g<1$ and $\lambda>0$.
Following standard arguments \cite{gogolin},
the ESR intensity vanishes identically for $\omega < u_1 k_\sigma$, 
with a power-law singularity approaching
 $\omega = u_1 k_\sigma$ from above,
\begin{equation} \label{I1}
I(\omega) \propto \theta(\omega - u_1 k_\sigma ) 
(\omega - u_1 k_\sigma)^{\xi_1^- + \xi_2^+ + \xi_2^- - 1 } \;.
\end{equation}
To see this we change integration variables  to 
$s=u_1 t-x$ and $s'= u_1 t+x$ in eq.~(\ref{esrfinal}).
In the new variables, the integrand exhibits branch cuts in the 
upper part of the $s$ and $s'$ complex plane but is analytic in the lower 
part.  This produces precisely the threshold behavior (\ref{I1}), 
where the exponent comes from a simple power-counting argument
related to a rescaling of $s'$. 
In order to investigate the behavior near $\omega=k_\sigma u_2$ it is
more convenient to use the change of variables $s=u_2 t -x$, $s'=u_2 t+x$. 
In the new variables power counting leads to the following power law
near 
$\omega=k_\sigma u_2 $:
\begin{equation}\label{I2}
I(\omega) \propto 
\left|\omega - u_2 k_\sigma \right|^{\xi_2^- + \xi_1^+ + \xi_1^- - 1 }
~.
\end{equation}
However, we cannot prove analytically that the intensity has a second 
threshold at this frequency, because the branch cuts of the integrand 
appear both in the upper and the lower half
 of the  $s$ and $s'$ planes. More generally, one can show
that because of $u_2 > u_1$ it is impossible to find a linear 
transformation, $s=a_{11} x + a_{12} t$ and
$s'=a_{21} x + a_{22} t$, such that the position of the branch cuts of
the integrand produces a $\theta (k_\sigma u_2-\omega)$ function. 
However,  numerical evaluation of eq.~(\ref{esrfinal})
indicates threshold behavior also at the upper edge.
Since in general $\xi^+_1 \neq \xi^+_2$, see
eq.~(\ref{exponents}), the ESR spectrum is markedly asymmetric
at both thresholds. The asymmetry is quite dramatic,
 because near $\omega = u_1 k_\sigma $, the 
intensity {\sl diverges} since the exponent is negative,  
whereas near $\omega = u_2 k_\sigma $ 
it {\sl vanishes} since the exponent is positive.

\begin{figure}
\onefigure{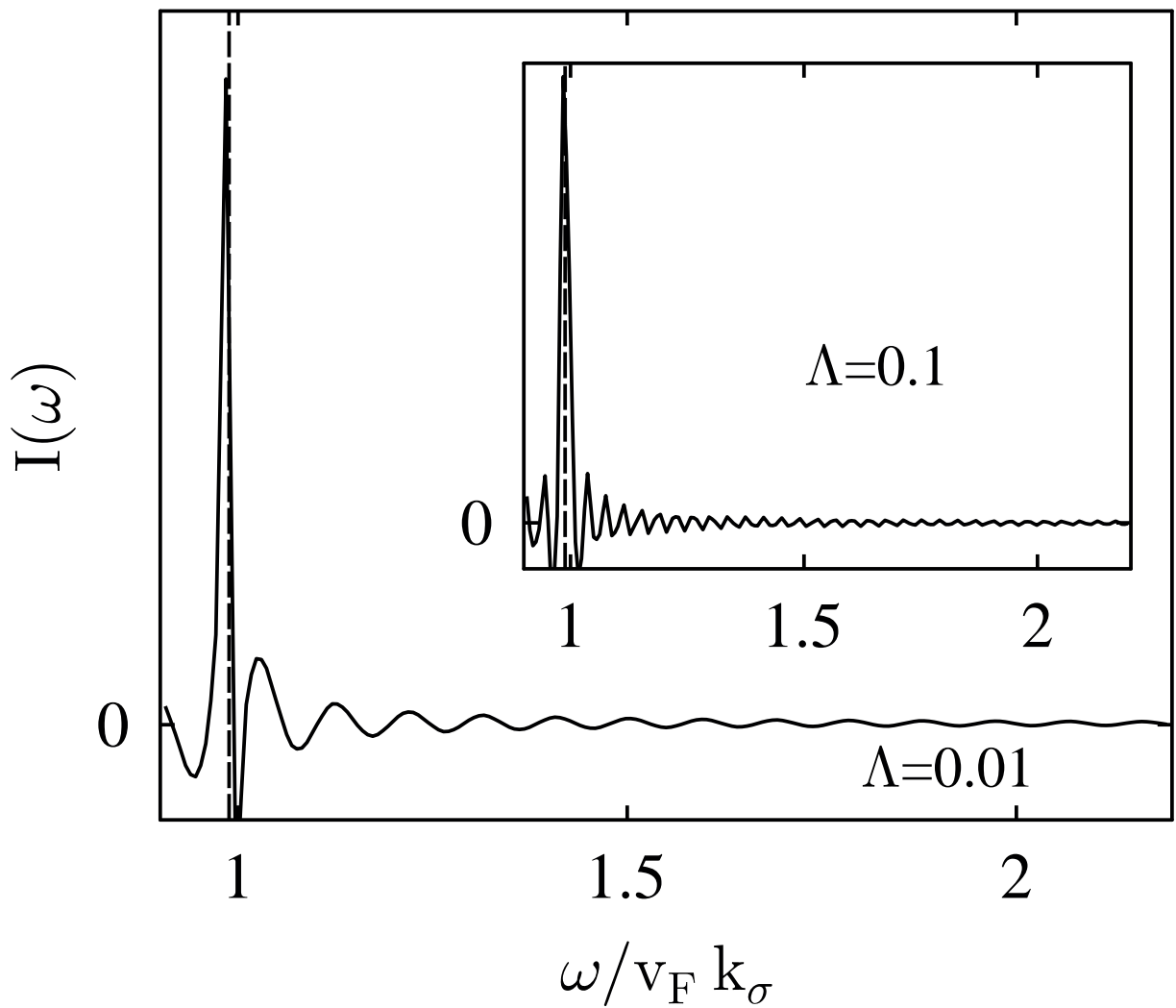}
\caption{ESR intensity (in arbitrary units) for $g=1/2$,  $\lambda=0.1$,
and two values of the cutoff $\Lambda$.
The threshold at $\omega=u_1 k_\sigma=0.988$ is indicated by the dashed line,
and the numerical result for $I(\omega)$ is given as the solid curve.
The ESR spectrum for $\Lambda\to 0$ terminates at $\omega=u_2 k_\sigma=2.011$. 
}
\label{fig1}
\end{figure}

\begin{figure}
\onefigure{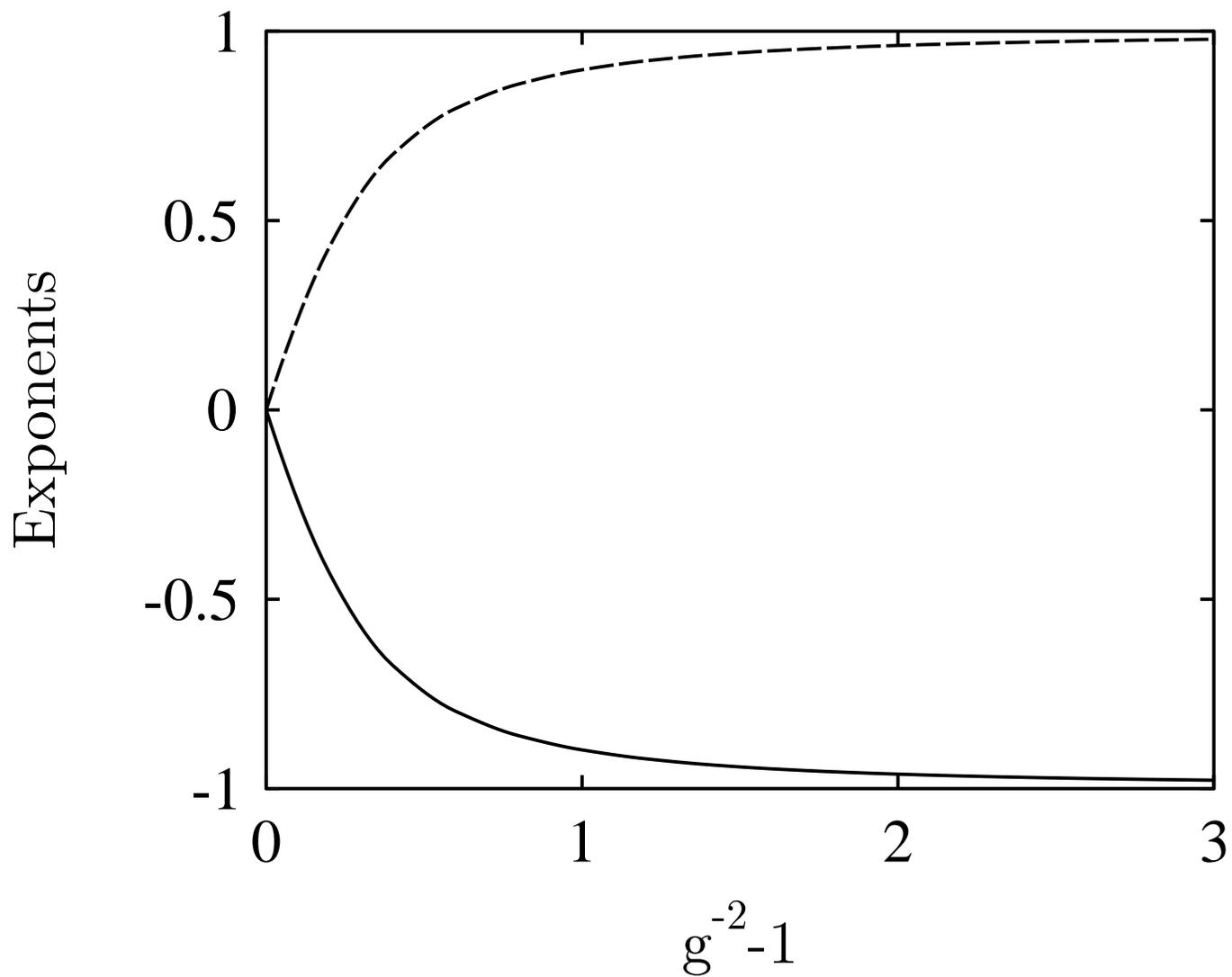}
\caption{Exponents near the lower threshold $\omega=u_1 k_\sigma$
(solid curve) and near the upper threshold $\omega=u_2 k_\sigma$ (dashed curve)
as function of the interaction strength for $\lambda=0.1$.}
\label{fig2}
\end{figure}

The full ESR spectrum can be obtained numerically 
by direct integration of eq.~(\ref{esrfinal}).
In the numerical integration, a finite cutoff $\Lambda$
has to be employed, which leads to several artefacts
that disappear for $\Lambda\to 0$, namely
(i) small oscillations,
(ii) finite but very small intensities outside the window
discussed above, and (iii) negative intensities.
Since the time discretization should resolve $\Lambda$, 
numerically one cannot choose arbitrarily small $\Lambda$,
and typical results for two values of $\Lambda$ are shown in fig.~\ref{fig1}.
Nevertheless, the data in fig.~\ref{fig1} are clearly consistent with 
the general behavior discussed above.
This behavior should be compared to both the non-interacting
case (where the power-law exponents are zero and a 
constant ESR intensity is found in the relevant window)
and to the $\lambda=0$ case (where one has a $\delta$-peak at the lower
edge). The interaction-strength-dependence of the exponents close to the
lower and upper threshold is shown in fig.~\ref{fig2}.
The exponents are of different sign,
but almost (yet not exactly) equal in magnitude. 

Using  $\xi^{\pm}= \xi^{\pm}_1 + \xi^{\pm}_2$ and
$\xi=\xi^+ +\xi^-$, we find for the integrated ESR spectrum 
\begin{equation}
\int \! \frac{d\omega}{2\pi} \, I(\omega) = 
\frac{2^{1-\xi}}{\pi \Gamma(\xi^+)}
\left[ \frac{\Gamma(-1 +\xi)}{\Lambda \Gamma(\xi^-)}
 + \frac{\Gamma(1-\xi)}{\Gamma(1-\xi^+)} (2k_\sigma)^{-1+\xi} 
\Lambda^{-2+\xi} \right]~.
\end{equation}
In the non-interacting limit, $\xi^- \rightarrow 0$ and 
$\xi^+ \rightarrow 2$, such that  the first term  
vanishes while the second term reproduces eq.~(\ref{intsp}).
Let us finally comment on the effect of thermal fluctuations ($T>0$).
Since the Hamiltonian is quadratic, these can be incorporated by
 substituting the $T=0$ propagators in eq.~(\ref{esrfinal}) according to
\[
\left(\frac{\pm i\Lambda}{x\mp u_it \pm i\Lambda}\right) \rightarrow
\left(\frac{\pm i\Lambda}{\frac{u_i}{\pi T}\sinh \frac{\pi T}{u_i}(x \mp u_it 
\pm 
i\Lambda)}\right) ~.
\]
In the non-interacting case, $I(\omega)$ can then be computed
exactly. This solution
demonstrates that thermal fluctuations wash out 
the thresholds, broaden the spectrum, and suppress the ESR intensity. 
In the interacting case, these features are also expected. In addition, 
scaling shows that the ESR intensity 
receives an overall power-law factor $T^{-2 + \xi}$.
The exponent is always positive, vanishes for $g=1$, and is
generally very small (for $\lambda=0.1$, it reaches a 
maximum value of $0.001$).  

To conclude, our study of the ESR spectrum resulting from
the combined presence of interactions and spin-orbit coupling
in 1D quantum wires reveals dramatic changes from
simple Fermi liquid modelling.  The appearance of
asymmetric power-law threshold behavior, with a  divergence 
at the lower edge and a vanishing intensity at the upper edge,
are manifestations of the breakdown of spin-charge separation
because of spin-orbit coupling.  Such features
should make an experimental study of our predictions
worthwhile and feasible in practice.

\acknowledgments
This work has been supported by the DFG under
the Gerhard-Hess program.

\end{document}